\documentclass[preprint,showpacs,preprintnumbers,amsmath,amssymb]{revtex4}

\usepackage{graphicx}
\usepackage{dcolumn}
\usepackage{bm}

\begin{document}

\preprint{}

\title{Near-threshold pion production with radioactive beams\\ at the Rare Isotope Accelerator}

\author{Bao-An Li}
\email{bali@astate.edu}
\affiliation{Department of Chemistry and
Physics, P.O. Box 419, Arkansas State University, State
University, Arkansas 72467-0419, USA}
\author{Gao-Chan Yong}
\author{Wei Zuo}
\affiliation{Institute of Modern Physics, Chinese Academy of
Science, Lanzhou 730000, P.R. China} \affiliation{Graduate School,
Chinese Academy of Science, Beijing 100039, P.R. China}
\date{\today}

\begin{abstract}
Using an isospin- and momentum-dependent transport model we study near-threshold pion production
in heavy-ion collisions induced by radioactive beams at the planned Rare Isotope Accelerator (RIA).
We revisit the question of probing the high density behavior of nuclear symmetry
energy $E_{sym}(\rho)$ using the $\pi^-/\pi^+$ ratio. It is found that both the total and
differential $\pi^-/\pi^+$ ratios remain sensitive to the $E_{sym}(\rho)$ when the
momentum-dependence of both the isoscalar and isovector potentials are consistently
taken into account. Moreover, the multiplicity and spectrum of $\pi^-$ mesons are found more
sensitive to the $E_{sym}(\rho)$ than those of $\pi^+$ mesons. Finally, effects of the Coulomb
potential on the pion spectra and $\pi^-/\pi^+$ ratio are also discussed.
\end{abstract}

\maketitle

\section{Introduction}
The isospin asymmetry of pions produced in nuclear reactions
was shown to be useful for extracting interesting information about the
structure of radioactive nuclei and neutron skins of heavy stable nuclei
within the Glauber model\cite{tel87,lom88,lihb91,ireview98,antoni}.
Within hadronic transport models the isospin asymmetry of pions was found useful also for studying
the equation of state (EOS) of isospin asymmetric nuclear matter\cite{uma,li03,gai04}.
In particular, the $\pi^-/\pi^+$ ratio was
proposed as a sensitive probe of the high density behavior of nuclear symmetry energy
$E_{sym}(\rho)$\cite{li03}. The latter is an important part of the nucleon specific
energy $E(\rho,\delta)$
\begin{equation}\label{ieos}
E(\rho ,\delta )=E(\rho ,\delta =0)+E_{\text{\textrm{sym}}}(\rho )\delta
^{2}+{\cal O}(\delta^4)
\end{equation}
in asymmetric matter of isospin asymmetry $\delta\equiv(\rho_{n}-\rho _{p})/(\rho _{p}+\rho _{n})$.
The density dependence of symmetry energy is very important for many interesting astrophysical
problems\cite{bethe,lat01,ibook01,bom1,steiner04}, the structure of radioactive
nuclei\cite{brown,horow,furn,stone} and heavy-ion reactions\cite{ireview98,ibook01}.
Unfortunately, the density dependence of symmetry energy, especially at supranormal densities,
is still very poorly known. Predictions based on various
many-body theories diverge widely at both low and high densities. In fact, even the sign of
the symmetry energy above $3\rho_0$ is still very uncertain\cite{bom1}.
Since nuclear reactions induced by high energy radioactive beams can produce transiently
dense neutron-rich matter, the fast fragmentation beams from RIA and the
new accelerator facility at GSI provide the first opportunity in terrestrial laboratories to explore
experimentally the ${\rm EOS}$ of dense neutron-rich
matter\cite{ireview98,li03,dan02,nsac02}. Crucial to the extraction of critical information
about the $E_{\rm sym}(\rho)$ is to compare experimental data with transport model
calculations. Based on isospin-dependent transport model calculations, several experimental
observables have been identified as promising probes of the $E_{sym}(\rho)$,
such as, the neutron/proton ratio \cite{li97}, isoscaling in nuclear
multifragmentation\cite{betty,tan01,bar02}, the neutron-proton
differential flow \cite{li00,gre03,sca,riz}, the neutron-proton correlation function\cite{chen03}
and the isobaric yield ratios of light clusters\cite{chen03b}. The first experimental constraint
on the density dependence of symmetry energy at subnormal densities was recently obtained
by analyzing the isospin diffusion data from heavy-ion reactions\cite{betty03,chen04b}.
While it is much more challenging to
constrain the symmetry energy at supranormal densities. In anticipation of coming experiments with
high energy radioactive beams at RIA and GSI, more theoretical studies identifying experimental
observables sensitive to the symmetry energy at supranormal densities are needed.

The single nucleon potential is an important input to the transport models. It includes
an isovector part (symmetry potential) and an isoscalar part, and both of them are momentum dependent
due to the non-locality of strong interactions and the Pauli exchange effects in
many-fermion systems. However, in all transport models for heavy-ion collisions the
momentum-dependence of the symmetry potential was seldom taken into account until very recently.
It was found that the momentum dependence of the symmetry potential affects many
experimental observables that were known to be sensitive to the symmetry
energy\cite{chen04b,lidas03,chen04a}. In this work we study near-threshold pion production at RIA
using a momentum and isospin dependent transport model\cite{lidas03}. We revisit the question
of whether the high density behavior of nuclear symmetry energy $E_{sym}(\rho)$ can be probed
using the $\pi^-/\pi^+$ ratio. We found that both the total and
differential $\pi^-/\pi^+$ ratios remain sensitive to the $E_{sym}(\rho)$ when the
momentum-dependence of both the isoscalar and isovector potentials are consistently
taken into account. Moreover, the multiplicity and spectrum of $\pi^-$ mesons are found more sensitive
to the $E_{sym}(\rho)$ than those of $\pi^+$ mesons.

\section{A brief summary of the isospin and momentum dependent transport model IBUU04}

In this work, we use the isospin and momentum dependent transport model
for heavy-ion collisions induced by neutron-rich nuclei\cite{lidas03}.
In the latest version of this model IBUU04 we use a single nucleon potential\cite{das03}
\begin{eqnarray}\label{mdi}
U(\rho,\delta,\vec p,\tau,x) &=& A_u(x)\frac{\rho_{\tau'}}{\rho_0}
+A_l(x)\frac{\rho_{\tau}}{\rho_0}\nonumber \\ &+&
B(\frac{\rho}{\rho_0})^{\sigma}(1-x\delta^2)-8\tau x\frac{B}{\sigma+1}\frac{\rho^{
\sigma-1}}{\rho_0^{\sigma}}\delta\rho_{\tau'} \nonumber \\
&+&\frac{2C_{\tau,\tau}}{\rho_0}
\int d^3p'\frac{f_{\tau}(\vec r,\vec p')}{1+(\vec p-\vec p')^2/\Lambda^2}\nonumber \\
&+&\frac{2C_{\tau,\tau'}}{\rho_0}
\int d^3p'\frac{f_{\tau'}(\vec r,\vec p')}{1+(\vec p-\vec p')^2/\Lambda^2}.
\end{eqnarray}
In the above $\tau=1/2$ ($-1/2$) for neutrons (protons) and $\tau\neq\tau'$; $\sigma=4/3$;
$f_{\tau}(\vec r,\vec p)$ is the phase space distribution
function at coordinate $\vec{r}$ and momentum $\vec{p}$.
The parameters $A_u(x), A_l(x), B, C_{\tau,\tau}, C_{\tau,\tau'}$ and $\Lambda$
were obtained by fitting the momentum-dependence of the
$U(\rho,\delta,\vec p,\tau,x)$ predicted by the Gogny Hartree-Fock and/or the Brueckner-Hartree-Fock
calculations, saturation properties of symmetric nuclear matter and the symmetry energy of 30 MeV
at normal nuclear matter density $\rho_0=0.16/fm^3$\cite{das03}. The compressibility of
symmetric nuclear matter $K_{0}$ is set to be 211 MeV.

The last two terms contain the momentum-dependence of the single particle potential.
The momentum dependence of the symmetry potential stems from the different interaction
strength parameters $C_{\tau,\tau'}$ and $C_{\tau,\tau}$
for a nucleon of isospin $\tau$ interacting, respectively, with unlike and like nucleons in
the background fields. More specifically, we use $C_{unlike}=-103.4$ MeV and $C_{like}=-11.7$ MeV.
One characteristic of the momentum dependence of the symmetry potential is the different effective
masses for neutrons and protons in isospin asymmetric nuclear matter. With the above potential,
we found that the neutron effective mass is higher than the proton effective mass and the
splitting between them increases with both the density and isospin asymmetry of the medium\cite{lidas03}.
Moreover, with the potential of eq.\ref{mdi} both the isoscalar and isovector potentials
at $\rho_0$ are in agreement with the corresponding nucleon optical potentials extracted
from nucleon-nucleus scattering data\cite{lidas03}.

The parameters $A_u(x)$ and $A_l(x)$ depend on the $x$ parameter according to
\begin{eqnarray}
A_u(x)&=&-95.98-x\frac{2B}{\sigma+1},\\
A_l(x)&=&-120.57+x\frac{2B}{\sigma+1}.
\end{eqnarray}
The parameter $x$ can be adjusted to mimic predictions on the density dependence of the
symmetry energy $E_{sym}(\rho)$ by microscopic and/or phenomenological many-body theories.
In this work we choose a range of the $x$ parameter from 1 to -2. With this choice the
density dependence of the symmetry energy samples a wide range of theoretical predictions
as shown in Fig.\ 1.

Other details of the IBUU04 model can be found in ref.\cite{lidas03}. The details of modeling pion
production can be found in our earlier publications, e.g., ref.\cite{li03,li91}.
Pion production was also studied previously by many other people
within transport models. However, none of the previous studies has taken into account the
momentum dependence of the symmetry potential. Only very recently the latter was recognized as an
important issue in connection with probing the high density behavior of symmetry energy
using the $\pi^-/\pi^+$ ratio.

\section{Results and discussions}
We study the reaction of $^{132}Sn+^{124}Sn$ at a beam energy of 400 MeV/nucleon
and an impact parameter of 1 fm as an example of high energy central reactions at RIA.
This particular reaction will be carried out using the fast fragmentation beam line at RIA.
The beam energy selected here is about the highest one to be available at RIA.
A Time Projection Chamber (TPC) has been proposed to study the EOS of isospin asymmetric
nuclear matter by detecting charged particles including pions\cite{tpc}.
In the following we discuss several features of
near-threshold pion production at RIA. The emphasis will be on examining their sensitivities
to the density dependence of symmetry energy.
\subsection{Formation of high density isospin asymmetric nuclear matter at RIA}
What are the maximum baryon density and isospin asymmetry that can be achieved in central heavy-ion
collisions at RIA? This is an interesting question relevant to the study of the EOS of
asymmetric nuclear matter. To answer this question we show in Fig.\ 2 the central
baryon density (upper window) and the average $(n/p)_{\rho\geq \rho_0}$ ratio (lower window)
of all regions with baryon densities higher than $\rho_0$. It is seen that the maximum baryon
density is about 2 times normal nuclear matter density. Moreover, the compression is rather
insensitive to the symmetry energy because the latter is relatively small compared to the
EOS of symmetric matter around this density. The high density phase lasts for about 15 fm/c from
5 to 20 fm/c for this reaction. It is interesting to see that the isospin asymmetry of the
high density region is quite sensitive to the symmetry energy. The soft (e.g., $x=1$) symmetry energy
leads to a significantly higher value of $(n/p)_{\rho\geq \rho_0}$ than the stiff one (e.g., $x=-2$).
This is consistent with the well-known isospin fractionation phenomenon.
Because of the $E_{sym}(\rho)\delta^2$ term in the EOS of asymmetric
nuclear matter, it is energetically more favorable to have a higher isospin asymmetry $\delta$
in the high density region with a softer symmetry energy functional $E_{sym}(\rho)$.
In the supranormal density region, as shown in Fig.\ 1, the symmetry
energy changes from being soft to stiff when the parameter $x$ varies from 1 to -2. Thus the value of
$(n/p)_{\rho\ge \rho_0}$ becomes lower as the parameter $x$ changes from 1 to -2. It is worth
mentioning that the initial value of the quantity $(n/p)_{\rho\ge \rho_0}$ is about 1.4 which is
less than the average n/p ratio of 1.56 of the reaction system. This is because of the neutron-skins
of the colliding nuclei, especially that of the projectile $^{132}Sn$. In the neutron-rich nuclei,
the n/p ratio on the low-density surface is much higher than that in their interior.
It is clearly seen that the dense region can become either neutron-richer or neutron-poorer
with respect to the initial state depending on the symmetry energy functional $E_{sym}(\rho)$ used.

\subsection{Dynamics of pion productions in heavy-ion collisions induced by radioactive beams}
To understand the dynamics of pion production and its dependence on the symmetry energy,
we show in Fig.\ 3 the multiplicity of $\pi^+$, $\pi^-$ and $\Delta(1232)$ as a function of time.
The multiplicity of $\Delta(1232)$ resonances shown in the figure includes all four charge states
while in the model we do treat and follow separately different charge states of the
$\Delta(1232)$ and $N^*(1440)$ resonances. At a beam energy of 400 MeV/nucleon which
is just about 100 MeV above the pion production threshold in nucleon-nucleon scatterings,
almost all pions are produced through the decay of $\Delta(1232)$ resonances. The contribution
due to $N^*$ resonances is negligible. By comparing Fig.\ 2
and Fig.\ 3 one can notice that most of the $\Delta$ resonances are produced in the high
density region. Pions from the decay of these resonances thus carry useful information about the
high density phase. It is interesting to see that the $\pi^-$ multiplicity depends more sensitively
on the symmetry energy. This feature is seen more clearly in Fig.\ 4 where the $\pi^-$ and $\pi^+$
multiplicities at the freeze-out are shown as a function of the $x$ parameter. While the $\pi^+$
multiplicity remains about the same the $\pi^-$ multiplicity increases by about 20\%
by varying the $x$ parameter from -2 to 1. The multiplicity of $\pi^-$ is about 2 to 3 times that of
$\pi^+$. This is because the $\pi^-$ mesons are mostly produced from neutron-neutron collisions.
The $\pi^-$ mesons are thus more sensitive to the isospin asymmetry of the reaction system and the
symmetry energy. In fact, assuming pions are all produced through $\Delta(1232)$ resonances in the
first chance nucleon-nucleon scatterings and neglecting the influence of subsequent
pion rescatterings and reabsorptions, the $\pi^-/\pi^+$ ratio is expected to scale
with the N/Z ratio of the participant region according to\cite{stock}
\begin{equation}
\pi^-/\pi^+=(5N^2+NZ)/(5Z^2+NZ)\approx (N/Z)^2.
\end{equation}
In reactions induced by neutron-rich nuclei one thus expects to see more $\pi^-$ than $\pi^+$ mesons.
For the reaction considered here, the average $(N/Z)^2$ of the reaction system is about 2.4.
The observed $\pi^-/\pi^+$ ratio is somewhat different from this value depending on the
symmetry energy used. It indicates the effects of the different $N/Z$ ratios of the participant regions
with the different $x$ parameters due to the isospin fractionation. Moreover, it also indicates
the importance of pion reabsorption and rescatterings which are both taken into account properly
in our transport model investigations.

Our finding that $\pi^-$ mesons are more sensitive to the symmetry energy has interesting implications for
the experimental studies. With the TPC, for instance, the $\pi^-$ mesons can be easily identified
because of their negative charges. While the $\pi^+$ mesons will be bent to the same direction in the
magnetic field as protons and are thus difficulty to be separated out cleanly. To investigate further the
possibility of exploring the symmetry energy using $\pi^-$ mesons alone, we show in Fig.\ 5 the
$\pi^-$ and $\pi^+$ kinetic energy spectra. It is seen that the $\pi^-$ spectra with different
$x$ parameters can indeed be clearly separated. It thus raises the interesting possibility of probing the
density dependence of symmetry energy by studying the $\pi^-$ spectrum.

\subsection{The $\pi^-/\pi^+$ ratio probe of the high density behavior of symmetry energy}
We now turn to the $\pi^-/\pi^+$ ratio as a probe of the high density behavior of symmetry energy.
The advantage of using the $\pi^-/\pi^+$ ratio over the $\pi^-$ spectrum itself is that the
ratio can reduce largely the systematic errors involved in the experiments.
Moreover, within the statistical model for pion production\cite{nature,bona}, the $\pi^-/\pi^+$ ratio is
proportional to ${\rm exp}\left[(\mu_n-\mu_p)/T\right]$,
where T is the temperature, $\mu_n$ and $\mu_p$ are the chemical potentials of neutrons and protons,
respectively. At modestly high temperatures ($T\geq 4$ MeV), the difference in the neutron and
proton chemical potentials can be written as\cite{thermal}
\begin{equation}\label{sta}
\mu_n-\mu_p=U^n_{asy}-U^p_{asy}-U_{Coulomb}+T\left[{\rm ln}\frac{\rho_n}{\rho_p}+\sum_m\frac{m+1}{m}B_m(\frac{\lambda_T^3}{2})^m(\rho^m_n-\rho^m_p)\right],
\end{equation}
where $U_{Coulomb}$ is the Coulomb potential for protons, $\lambda_T$ is the thermal wavelength of
a nucleon and $B'_m$s are the inversion coefficients of the Fermi distribution function\cite{thermal}.
The difference in neutron and proton symmetry
potentials $U^n_{asy}-U^p_{asy}\approx 2U_{asy}\delta$, where the function
$U_{asy}$ is the strength of the symmetry potential, depends on the $x$ parameter,
the density $\rho$ and the nucleon momentum. It can be estimated readily
from the single nucleon potential of eq.\ref{mdi}.
It is also seen that the kinetic part of the difference $\mu_n-\mu_p$ relates directly to the
isospin asymmetry $\rho_n/\rho_p$ or $\rho_n-\rho_p$. Thus from the statistical point of view
the $\pi^-/\pi^+$ ratio is theoretically a good probe of the symmetry energy. However,
to verify this expectation and study more realistically the $\pi^-/\pi^+$ ratio one has
to rely on the transport models.

Shown in Fig.\ 6 is the quantity $(\pi^-/\pi^+)_{like}$ as a function of time. Taking into account the
dynamics of resonance production and decays we define
\begin{equation}
(\pi^-/\pi^+)_{like}\equiv \frac{\pi^-+\Delta^-+\frac{1}{3}\Delta^0}
{\pi^++\Delta^{++}+\frac{1}{3}\Delta^+}.
\end{equation}
This ratio naturally becomes the final $\pi^-/\pi^+$ ratio after all resonances have decayed.
First, it is seen that the $(\pi^-/\pi^+)_{like}$ ratio reaches a very high value in the early
stage of the reaction. This is due to the abundant neutron-neutron scatterings when the two neutron-skins
start overlapping at the beginning of the reaction. Secondly, the $(\pi^-/\pi^+)_{like}$
ratio saturates after about 25 fm/c indicating that a chemical freeze-out stage has been reached. Finally,
the sensitivity to the symmetry energy is clearly shown in the final $\pi^-/\pi^+$ ratio.
One can notice that the sensitivity of the $\pi^-/\pi^+$ ratio to the $x$ parameter
is quantitatively about the same as the $\pi^-$ multiplicity shown in Fig.\ 4.
Shown in Fig. 7 and Fig. 8 are the differential $\pi^-/\pi^+$ ratios versus the
kinetic energy and transverse momentum, respectively. In the low energy ($E_{kin}\leq 120$ MeV)
or transverse momentum ($p_t \leq 200$ MeV/c) region the $\pi^-/\pi^+$ ratio is clearly separable with
the $x$ parameter varying from 1 to -2. It indicates that it is sufficient to
measure accurately the low energy (transverse momentum) pions, instead of the whole spectrum, in order
to constrain the density dependence of the symmetry energy.

\subsection{Coulomb effects on the $\pi-/\pi+$ ratio}
First of all, it is necessary to mention that the Coulomb effects on $\pi^-/\pi^+$ ratio is
well known, in particular from experiments at the BEVALAC and SIS/GSI,
see, e.g., ref. \cite{coulomb} and references therein. A number of models
have been used in analysing the $\pi^-/\pi^+$ ratio in heavy-ion reactions from
low to ultra-relativistic energies. However, in previous studies the effects of the
symmetry energy was generally neglected.
From eq. \ref{sta} it is seen that the  $\pi^-/\pi^+$ ratio depends on both the Coulomb and
symmetry potentials. Here we are interested in understanding the relative effects of the
symmetry and Coulomb potentials. As an example, we compare in Fig.\ 9 the multiplicities
of $\pi^-$ and $\pi^+$ mesons as a function of transverse momentum calculated with and without the
Coulomb potential with the parameter $x=0$. It is seen that the Coulomb potential is to
shift $\pi^-$ ($\pi^+$) to lower (higher) $p_t$ as one expects. The $\pi^-/\pi^+$ ratio
with and without the Coulomb potential is shown in Fig.\ 10. It is seen that the $\pi^-/\pi^+$
ratio is quite flat without the Coulomb potential. The Coulomb effect increases the $\pi^-/\pi^+$
ratio at $p_t=0$ for about 30\% while suppresses it at higher $p_t$.
What is the relative effect of the symmetry potential with respect to that of the Coulomb potential?
The answer to this question can be obtained from inspecting Fig. 7 and Fig. 8.
With different $x$ parameters, the reaction dynamics is about the same as indicated by
the almost identical evolution of central density shown in Fig.\ 2. Thus the Coulomb
effect should also be about the same in calculations with the different $x$ parameters.
From the variation of the $\pi^-/\pi^+$ ratio by varying the $x$ parameter as shown in
Figs. 7 and 8, we can conclude that the Coulomb effect is stronger than the symmetry potential.
Of course, we should stress that the symmetry potential has an indirect effect on
the $\pi^-/\pi^+$ ratio through its interactions
on nucleons, while the Coulomb potential acts directly on charged pions. Since the
Coulomb effect is well-known, our results discussed above indicate that it is very
promising to extract useful information about the symmetry potential from
studying the $\pi^-/\pi^+$ ratio.

\section{Summary}
In summary, using an isospin- and momentum-dependent transport model
we have studied near-threshold pion production in heavy-ion collisions induced
by radioactive beams at the planned Rare Isotope Accelerator (RIA).
We studied properties of the high density matter formed in high energy
central reactions at RIA. It is found that the isospin asymmetry of the
high density hadronic matter is very sensitive to the symmetry energy used.
With the soft symmetry energy the n/p ratio of the high density region can be
significantly higher than that of the reaction system due to the isospin fractionation.
We also examined the multiplicities and spectra of both $\pi^-$ and $\pi^+$ mesons.
We found that the $\pi^-$ mesons carry more sensitive information about the symmetry energy
than the $\pi^+$ mesons. Both the kinetic energy and transverse momentum spectra of $\pi^-$ mesons
can be useful for studying the symmetry energy at high densities.
Moreover, we revisited the question of probing the high density behavior of nuclear symmetry
energy $E_{sym}(\rho)$ using the $\pi^-/\pi^+$ ratio. It was found that both the total and
differential $\pi^-/\pi^+$ ratios remain sensitive to the $E_{sym}(\rho)$ when the
momentum-dependence of both the isoscalar and isovector potentials are consistently
taken into account. Furthermore, we found that the effects of the Coulomb potential on the
$\pi^-/\pi^+$ ratio are important. In fact, the Coulomb effect is stronger than that of the
symmetry potential. Nevertheless, taking into account the Coulomb effect and the momentum
dependence of the symmetry potential the $\pi^-/\pi^+$ ratio, especially in the low energy or
transverse momentum region, remains very sensitive to the variation of the symmetry energy.
Our study thus confirms that the isospin asymmetry of pions from heavy-ion reactions induced
by radioactive beams is a very promising tool for studying the EOS of dense neutron-rich matter,
especially the high density behavior of symmetry energy. Our findings here are expected to be
useful for designing the TPC and planning experiments at RIA.

\section{Acknowledgments}
We would like to thank Scott Pratt,
Jianye Liu and Xiguo Lee for helpful discussions. The work of B.A.
Li is supported in part by the National Science Foundation of the
United States under grant No. PHYS-0243571 and PHYS0354572. The
work of G.C. Yong and W. Zuo is supported in part by the Chinese
Academy of Science Knowledge Innovation Project (KECK2-SW-N02),
Major State Basic Research Development Program (G2000077400), the
National Natural Science Foundation of China (10235030) and the
Important Pare-Research Project (2002CAB00200) of the Chinese
Ministry of Science and Technology.

\newpage
\begin{figure}
\includegraphics[scale=0.55,angle=-90]{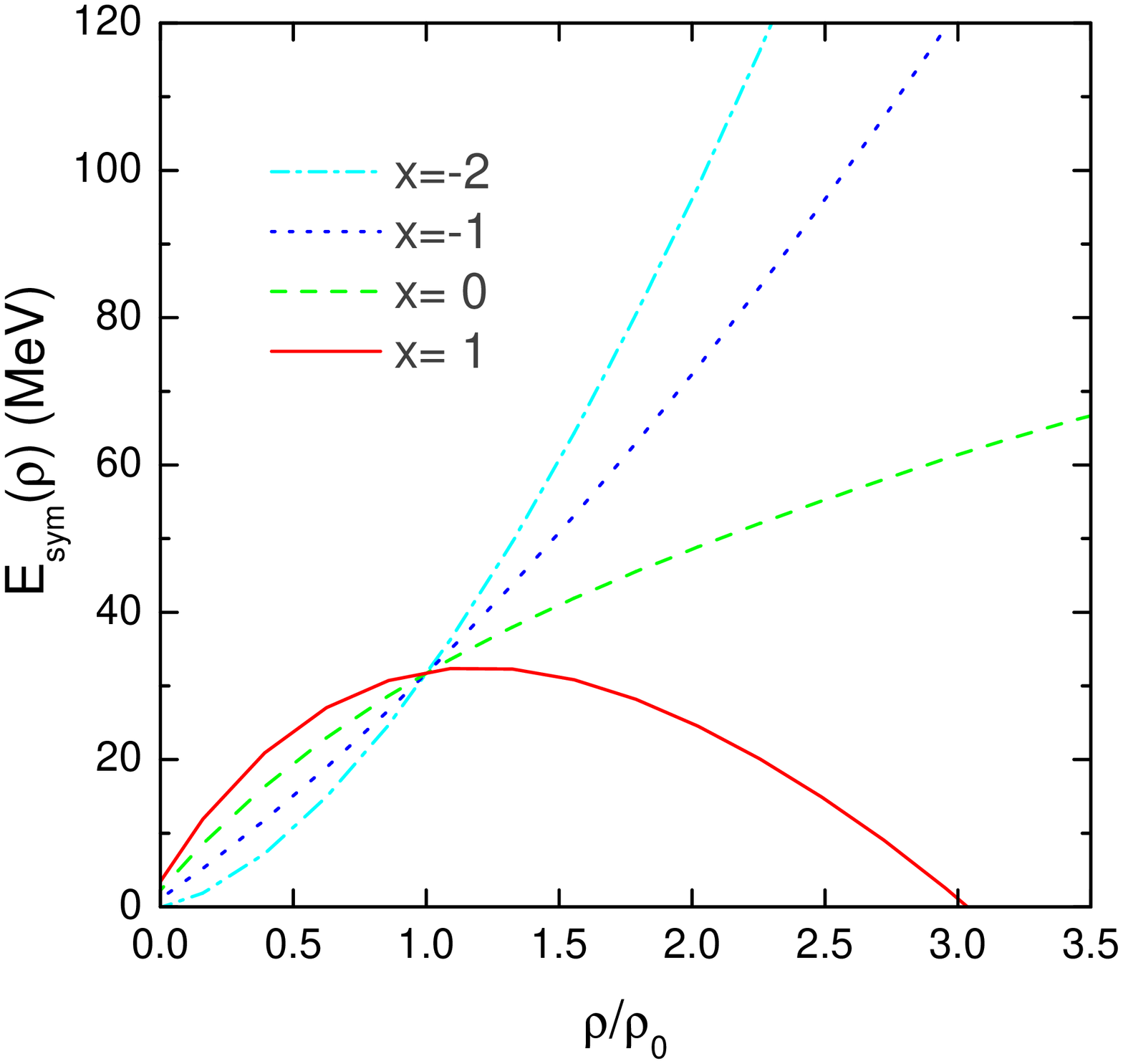}
\vspace{1.cm} \caption{{\protect\small (Color on line) Nuclear
symmetry energy as a function of density with different $x$
parameter.}} \label{esym}
\end{figure}
\begin{figure}
\includegraphics[scale=0.55,angle=-90]{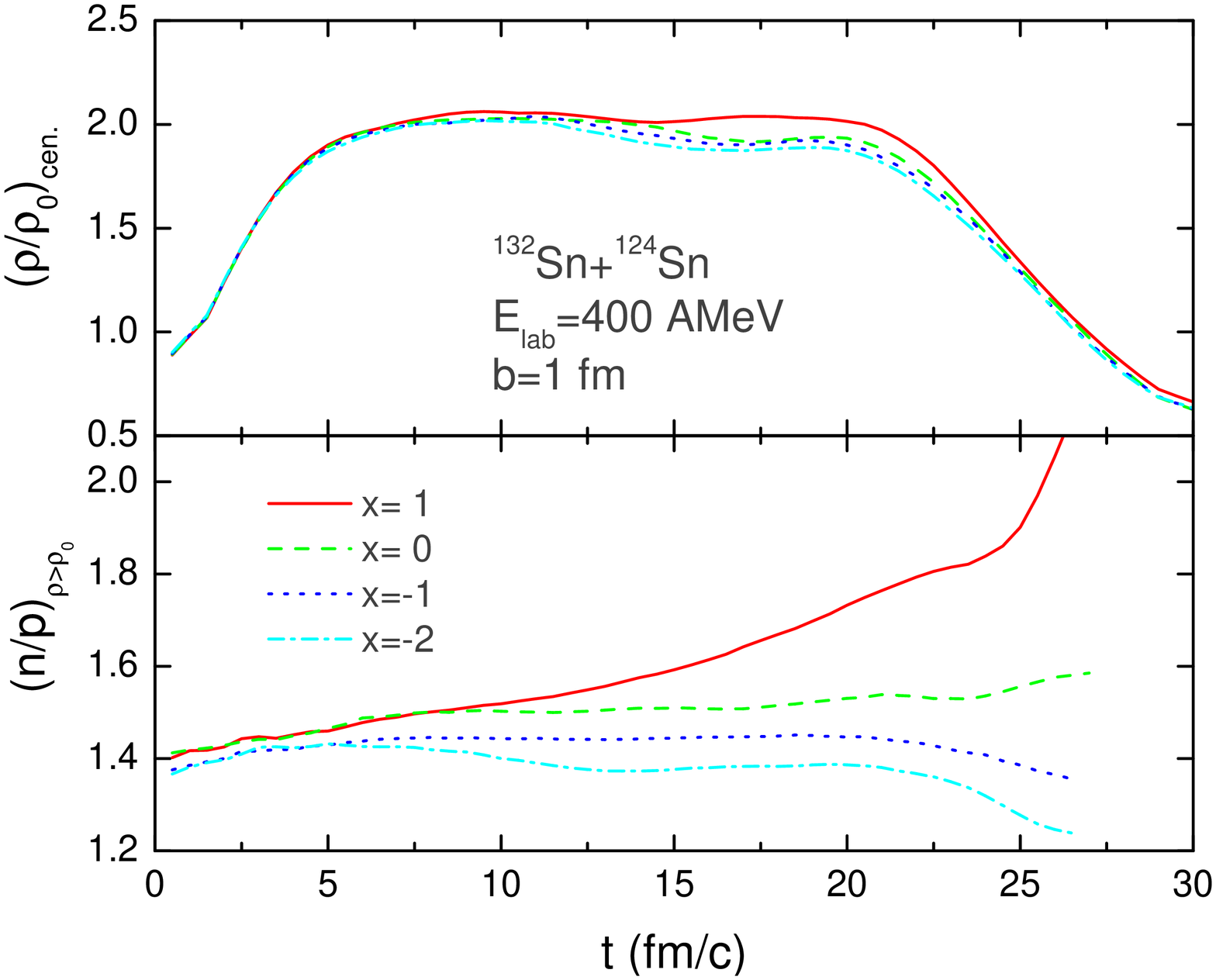}
\vspace{1.cm} \caption{{\protect\small (Color on line) Central
baryon density (upper window) and isospin asymmetry (lower window)
of high density region for the reaction of $^{132}Sn+^{124}Sn$ at
a beam energy of 400 MeV/nucleon and an impact parameter of 1
fm.}} \label{density}
\end{figure}

\begin{figure}
\includegraphics[scale=0.55,angle=-90]{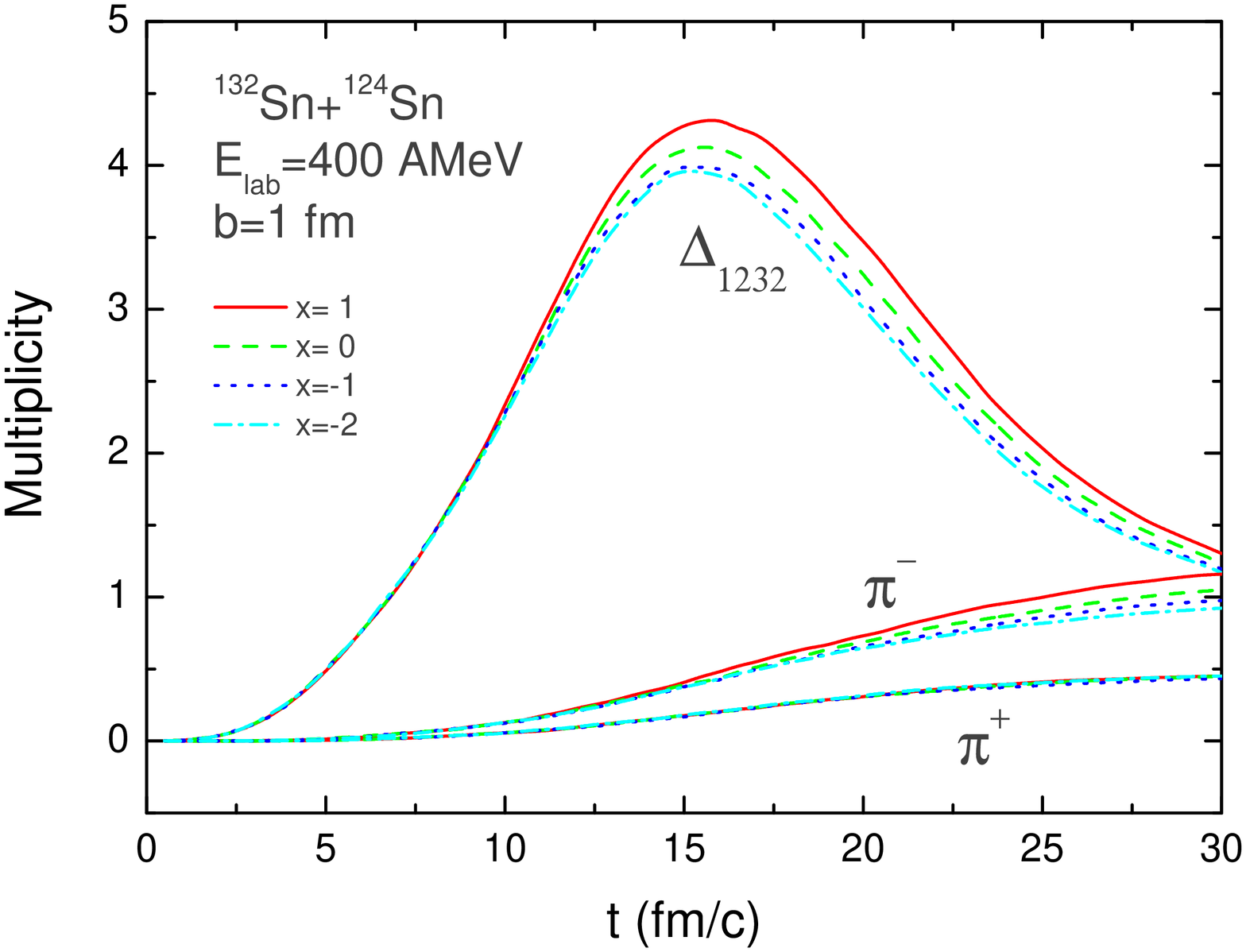}
\vspace{1.cm} \caption{{\protect\small (Color on line) Evolution
of the pion and $\Delta(1232)$ multiplicity in the reaction of
$^{132}Sn+^{124}Sn$ at a beam energy of 400 MeV/nucleon and an
impact parameter of 1 fm.}} \label{multiplicity}
\end{figure}

\begin{figure}
\includegraphics[scale=0.55,angle=-90]{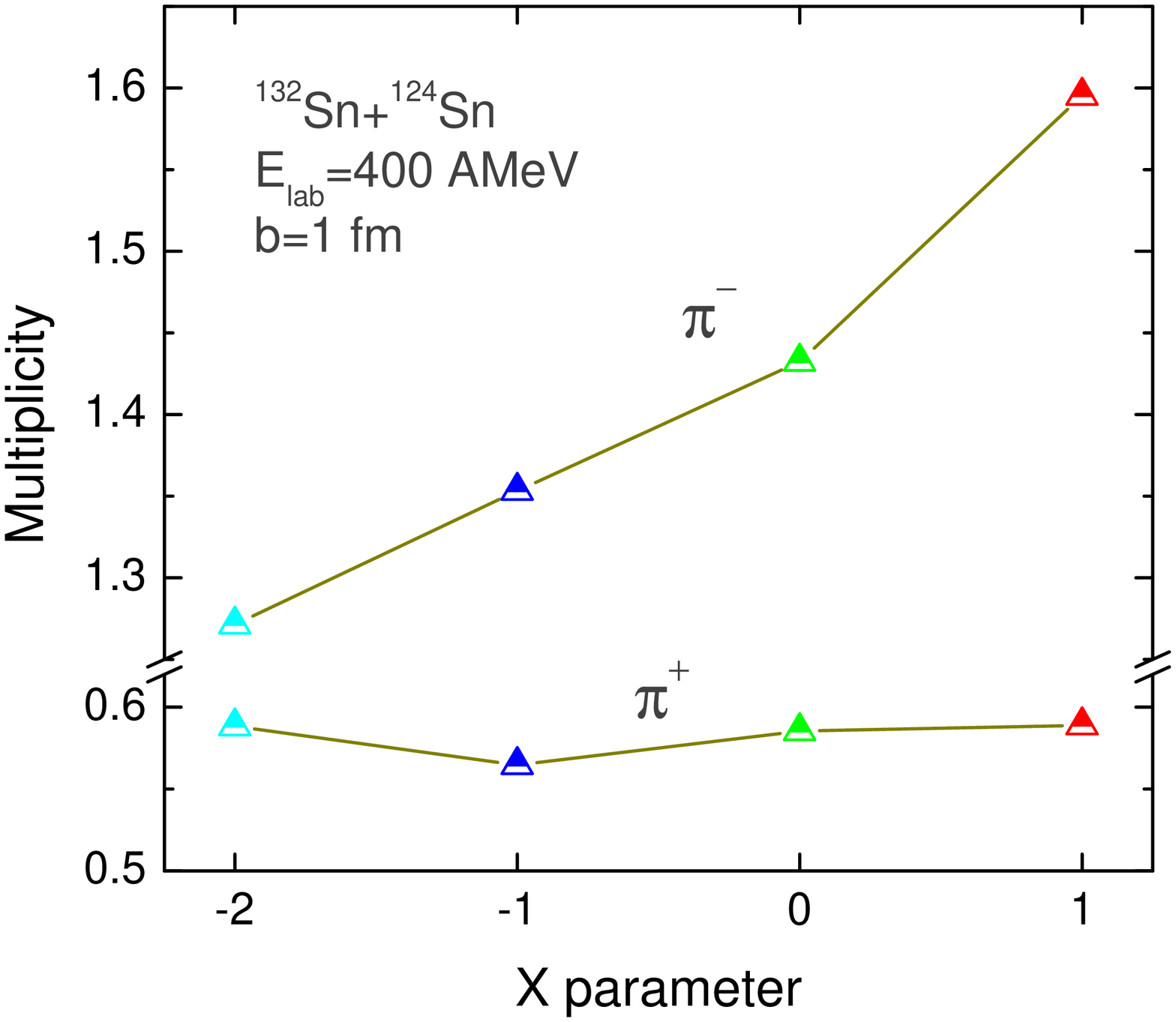}
\vspace{1.cm} \caption{{\protect\small (Color on line) The average
multiplicity of $\pi^+$ and $\pi^-$ as a function of the $x$
parameter for the reaction of $^{132}Sn+^{124}Sn$ at a beam energy
of 400 MeV/nucleon and an impact parameter of 1 fm.}}
\label{xmultiplicity}
\end{figure}

\begin{figure}
\includegraphics[scale=0.55,angle=-90]{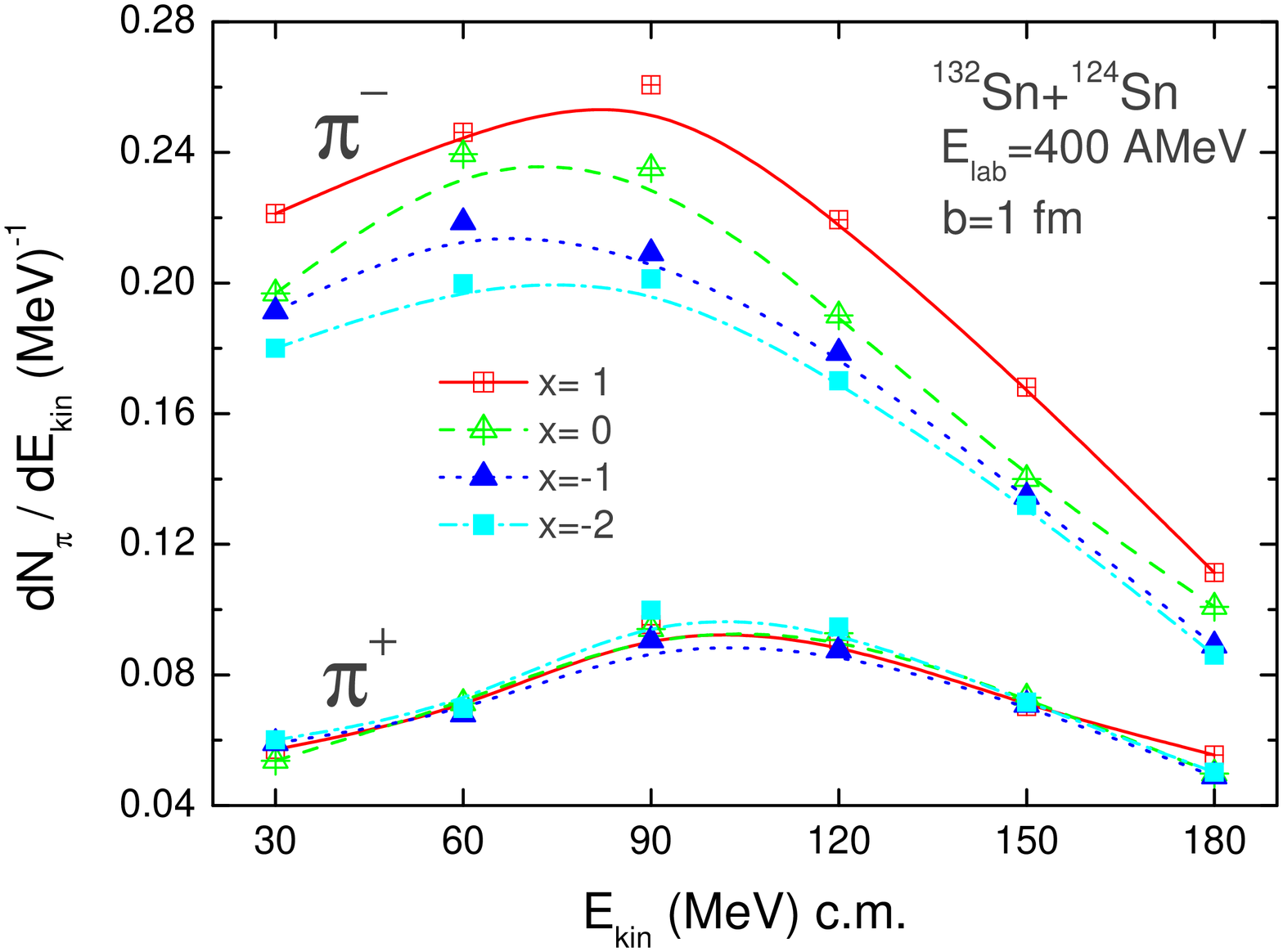}
\vspace{1.cm} \caption{{\protect\small (Color on line) The kinetic
energy spectra of $\pi^-$ and $\pi^+$ in the reaction of
$^{132}Sn+^{124}Sn$ at a beam energy of 400 MeV/nucleon and an
impact parameter of 1 fm.}} \label{spectra}
\end{figure}

\begin{figure}
\includegraphics[scale=0.55,angle=-90]{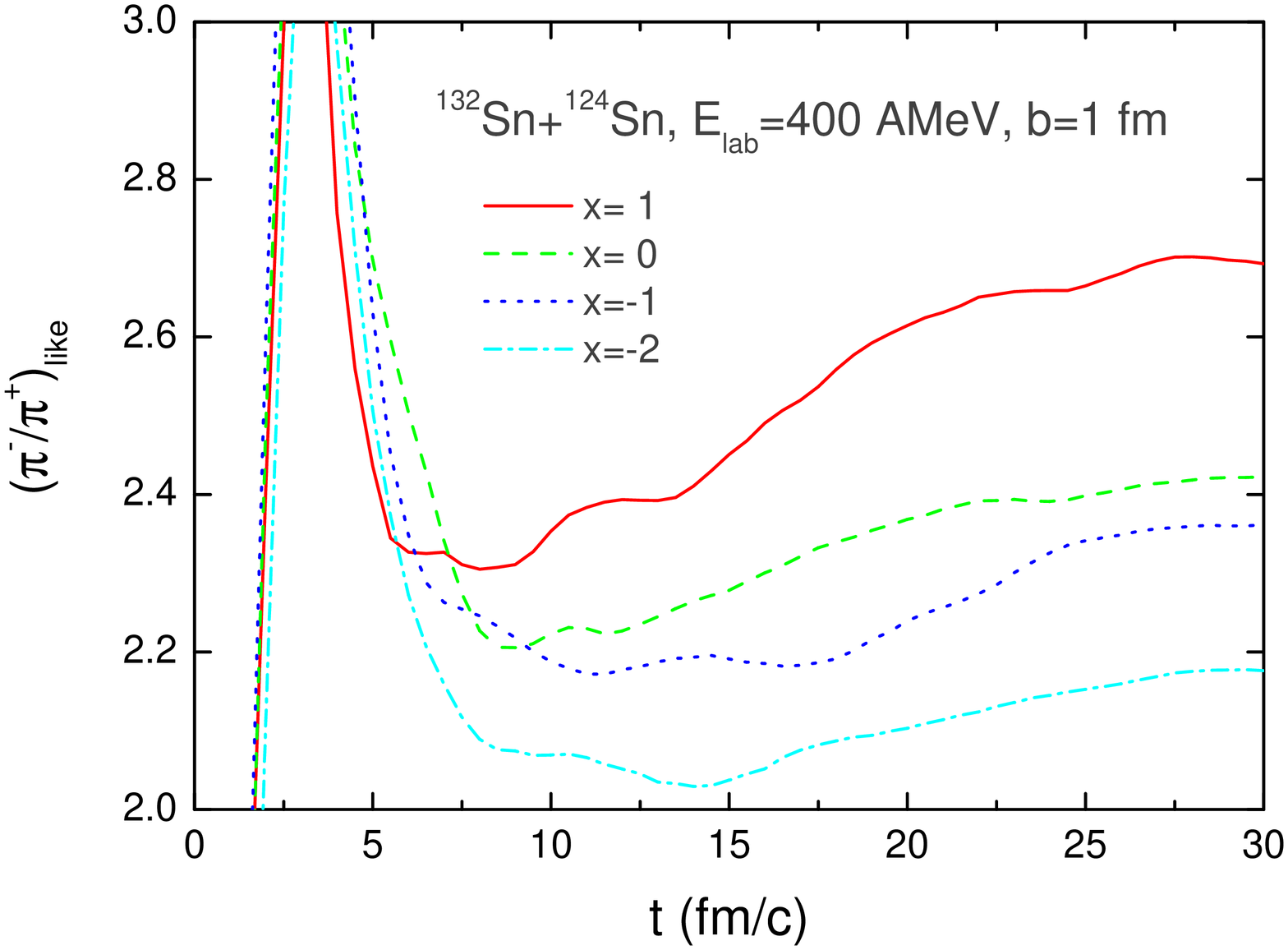}
\vspace{1.cm} \caption{{\protect\small (Color on line) Evolution
of the $\pi^-/\pi^+$ ratio in the reaction of $^{132}Sn+^{124}Sn$
at a beam energy of 400 MeV/nucleon and an impact parameter of 1
fm.}} \label{ratio}
\end{figure}

\begin{figure}
\includegraphics[scale=0.55,angle=-90]{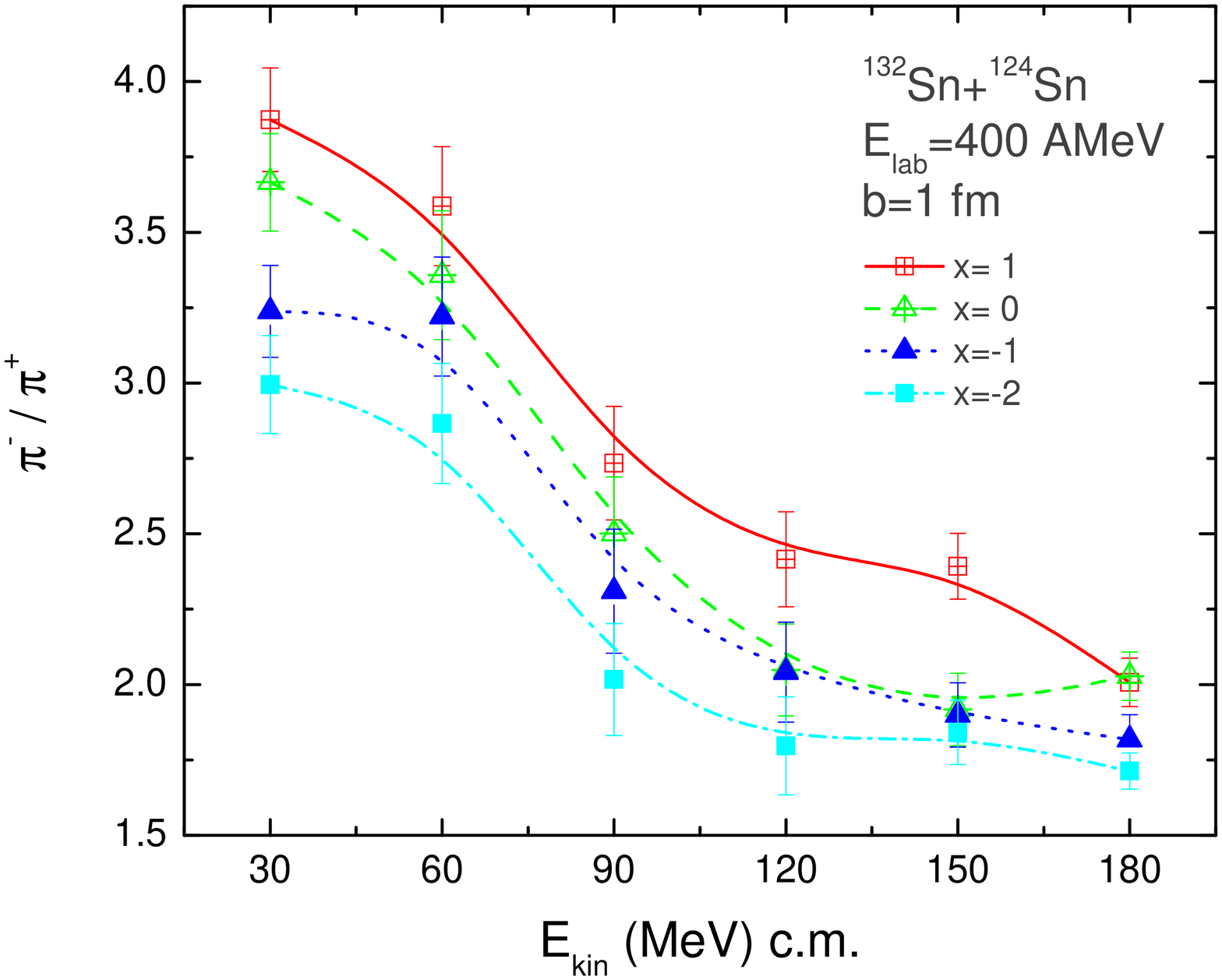}
\vspace{1.cm} \caption{{\protect\small (Color on line) The
$\pi^-/\pi^+$ ratio as a function of pion kinetic energy in the
reaction of $^{132}Sn+^{124}Sn$ at a beam energy of 400
MeV/nucleon and an impact parameter of 1 fm.}} \label{eratio}
\end{figure}

\begin{figure}
\includegraphics[scale=0.55,angle=-90]{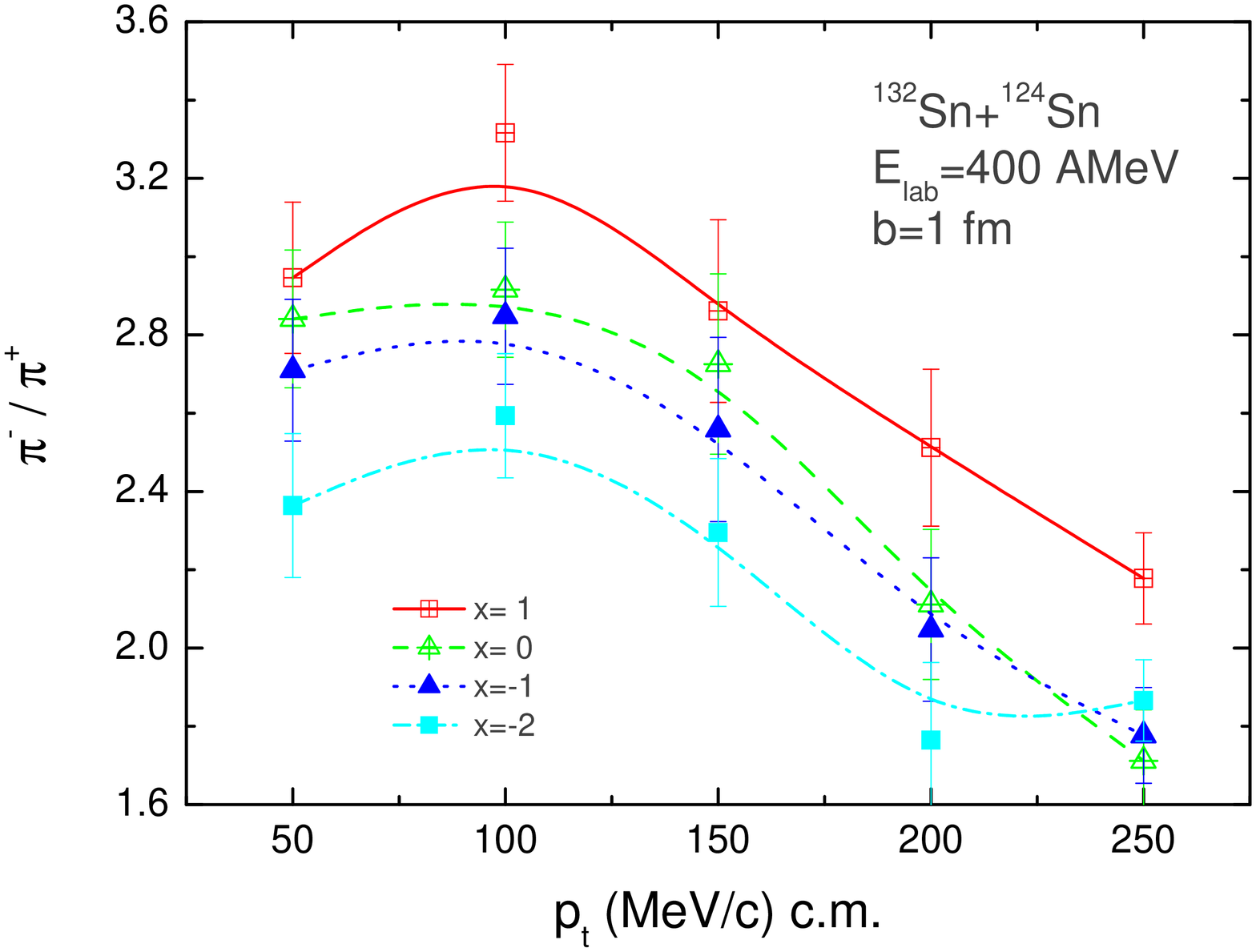}
\vspace{1.cm} \caption{{\protect\small (Color on line) The
$\pi^-/\pi^+$ ratio as a function of transverse momentum in the
reaction of $^{132}Sn+^{124}Sn$ at a beam energy of 400
MeV/nucleon and an impact parameter of 1 fm.}} \label{kratio}
\end{figure}

\begin{figure}
\includegraphics[scale=0.55,angle=-90]{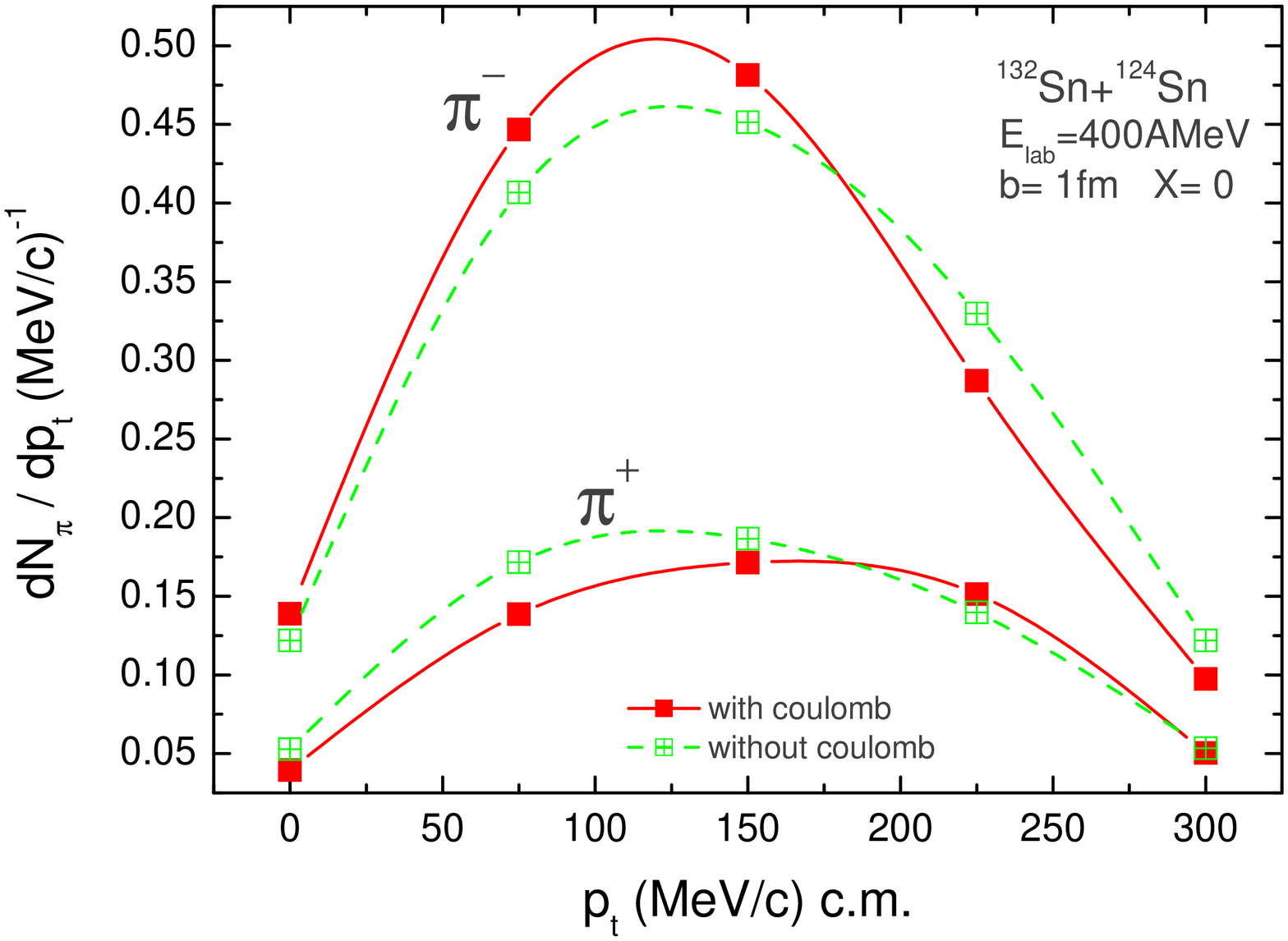}
\vspace{1.cm} \caption{{\protect\small (Color on line) $\pi^-$ and
$\pi^+$ transverse momentum spectra calculated with and without
the Coulomb potential for the reaction of $^{132}Sn+^{124}Sn$ at a
beam energy of 400 MeV/nucleon and an impact parameter of 1 fm.}}
\label{cspectra}
\end{figure}

\begin{figure}
\includegraphics[scale=0.55,angle=-90]{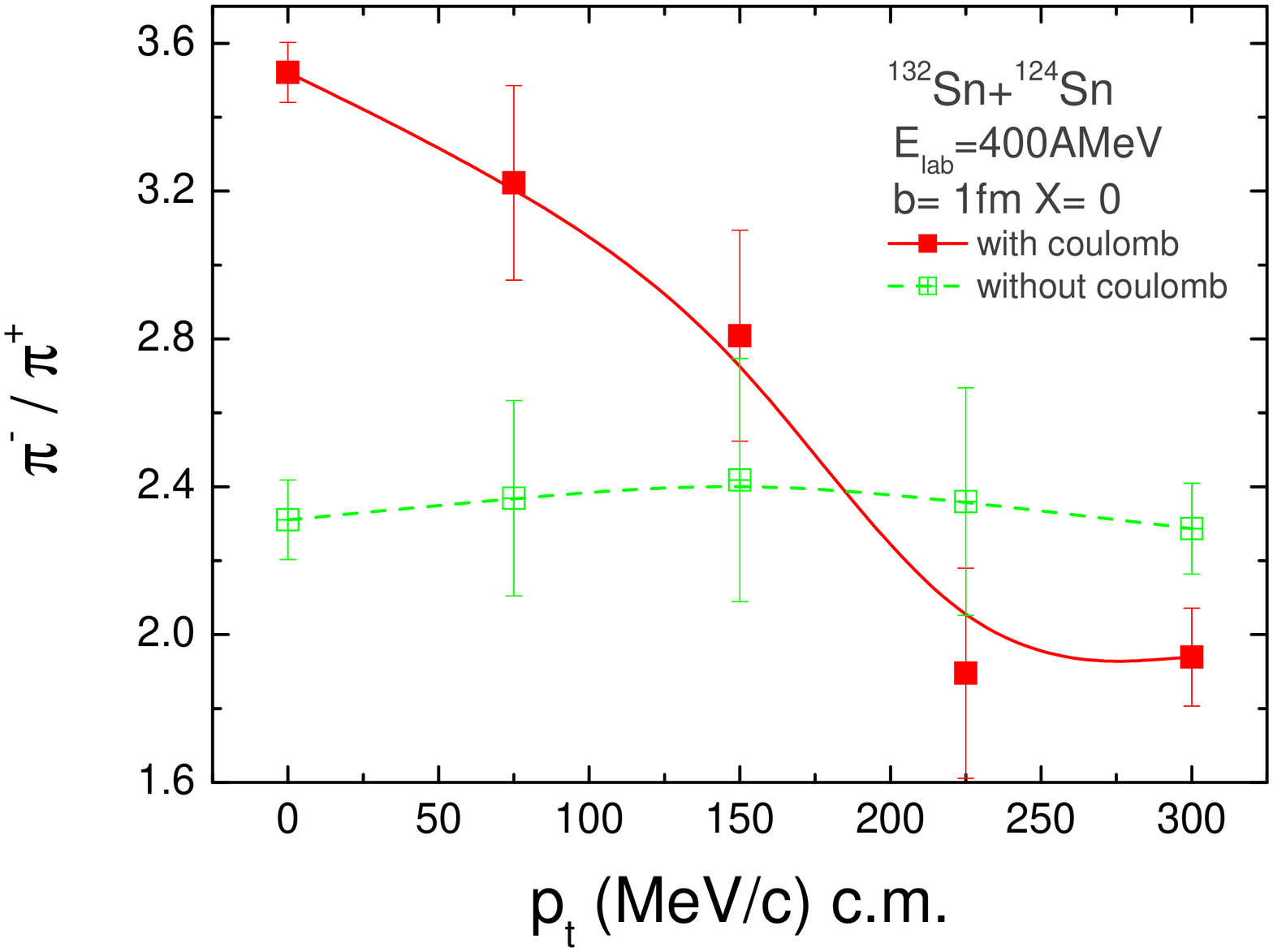}
\vspace{1.cm} \caption{{\protect\small (Color on line)
$\pi^-/\pi^+$ ratios as a function of transverse momentum
calculated with and without the Coulomb potential in the reaction
of $^{132}Sn+^{124}Sn$ at a beam energy of 400 MeV/nucleon and an
impact parameter of 1 fm.}} \label{cratio}
\end{figure}

\end{document}